\documentclass[aps,prl,twocolumn,preprintnumbers,amsmath,amssymb,superscriptaddress,10pt]{revtex4-2}
\usepackage{amsmath,graphicx}
\usepackage[utf8]{inputenc}
\usepackage[T1]{fontenc}
\usepackage{xcolor}
\usepackage{textcomp}
\usepackage{bm}

\usepackage{physics}
\usepackage{amsmath}
\usepackage{tikz}
\usetikzlibrary{arrows.meta}
\usepackage{mathdots}
\usepackage{yhmath}
\usepackage{cancel}
\usepackage{color}
\usepackage{array}
\usepackage{multirow}
\usepackage{amssymb}
\usepackage{gensymb}
\usepackage{tabularx}
\usepackage{extarrows}
\usepackage{booktabs}
\usetikzlibrary{fadings}
\usetikzlibrary{patterns}
\usetikzlibrary{shadows.blur}
\usetikzlibrary{shapes}

\usepackage{color}
\definecolor{LinkColor}{rgb}{0.75,0.0,0.2}

\usepackage{hyperref}
\hypersetup{
	pdfauthor={good guys},
	pdftitle={good title},
	colorlinks=true,
	citecolor=LinkColor,
	linkcolor=LinkColor,
	urlcolor=LinkColor,
}

\usepackage{listings}
\definecolor{lightgray}{gray}{1}

\usepackage{theorem}

\usepackage{tikz}

\newcommand{\nc}{\newcommand}
\nc{\braoprket}[3]{\langle#1|#2|#3\rangle}
\nc{\opn}[1]{\operatorname{#1}}
\nc{\avg}[1]{\langle#1\rangle}
\nc{\ketbrasame}[1]{|#1\rangle\!\langle#1|}
\nc{\swap}{\opn{SWAP}}
\nc{\E}{\mathbb{E}}
\nc{\Var}{\opn{Var}}
\nc{\dg}{\dagger}

\usepackage[normalem]{ulem}

\begin{document}

\title{Critical Topological Photonics in Synthetic Dimensions}

\author{Mingyuan Chen} 
\affiliation{Eastern Institute of Technology, Ningbo 315200, China}
\affiliation{Quantum Information Physics Theory Research Team, Center for Quantum Computing, RIKEN, Wakoshi, Saitama, 351-0198, Japan}

\author{Yuwei Jing}
\affiliation{College of Engineering and Applied Sciences, National Laboratory of Solid State Microstructures, and Collaborative Innovation Center of Advanced Microstructures, Nanjing University, Nanjing 210023, China}

\author{Franco Nori}
\email{fnori@riken.jp}
\affiliation{Quantum Information Physics Theory Research Team, Center for Quantum Computing, RIKEN, Wakoshi, Saitama, 351-0198, Japan}

\author{Xue-Jia Yu}
\email{xuejiayu@eitech.edu.cn}
\affiliation{Eastern Institute of Technology, Ningbo 315200, China}

\date{\today}

\begin{abstract}
Topological states and criticality have long been regarded as incompatible ingredients: the former requires a finite spectral gap, whereas the latter demands its closure. Guided by this view, topological photonics has focused almost exclusively on gapped phases, treating gap-closing transitions as mere phase boundaries. 
In this work, 
we propose a class of topological states in which topology coexists with criticality in experimentally accessible synthetic-frequency photonic platforms. In a one-dimensional (1D) synthetic lattice, we identify such critical topological photonic states through midgap degeneracies in the single-particle entanglement spectrum, and uncover a topology-enforced multicritical point that reorganizes the topology of neighboring critical states. We further extend this framework to two dimensions (2D). 
Our work provides an experimentally accessible route to critical topological photonics, and may also inspire novel applications such as critical topological sensing.
\end{abstract}

\maketitle

\emph{Introduction.}---Topological phases of matter are conventionally understood as gapped phenomena: quantized topological invariants such as winding and Chern numbers are defined on gapped bulk bands, and the bulk-boundary correspondence guarantees robust edge states precisely because the bulk spectral gap prevents their hybridization with the continuum~\cite{HasanKane2010,QiZhang2011,ChiuTeo2016}. Over the past decade, this gap-based framework has been extended from electronic systems to photonics, leading to the concept of photonic topological insulators~\cite{Ozawa2019,Lu2014,Leykam2026,Nori2015Science,PhysRevLett.128.203602,PhysRevLett.128.203602,khanikaev2017two,khanikaev2013photonic,PhysRevLett.134.223806,PhysRevLett.120.250501,bliokh2019topological,PhysRevB.102.045129,PhysRevB.102.134213,khanikaev2024topological,vakulenko2023adiabatic,leefmans2022topological} and enabling a broad range of topological photonic devices, including robust waveguides, topological lasers, and disorder-tolerant light-routing architectures~\cite{amelio2020theory,harari2018topological,bandres2018topological,yang2022topological,lu2021topological,PhysRevLett.134.203803,leefmans2024topological,kumar2025emission,barik2018topological,price2022roadmap}. Criticality, by contrast, is marked by a vanishing bulk spectral gap and is classified into different universality classes determined by a set of critical exponents within the Landau–Ginzburg paradigm~\cite{Sachdev2011,Cardy1996}. Consequently, these two notions have long been regarded as mutually exclusive, giving rise to the long-standing belief that the existence of topological states requires a bulk spectral gap and that topological properties are necessarily destroyed once the bulk band gap closes.

Recent progress has overturned this dichotomy through the discovery of topological physics in quantum critical systems, where robust topological edge modes coexist with a gapless bulk~\cite{Scaffidi2017,Verresen2018,Verresen2021,Yu2022PRL,YuXuLin2026,PhysRevB.103.L100207,verresen2020,PhysRevLett.129.210601,PhysRevB.106.144436,PhysRevLett.133.026601,cv5q-8t25,nj3d-8g9s,PhysRevLett.134.116602,PhysRevB.97.165114}. This conceptual breakthrough has given rise to the notion of critical topological states and attracted significant recent attention, as it establishes a new paradigm in condensed matter and statistical physics: (i) topological states are not restricted to gapped systems and can exhibit richer topological phenomena unique to gapless regimes; and (ii) topology can further enrich universality classes even when the critical exponents coincide~\cite{YuXuLin2026}. Despite these theoretical advances, critical topological states remain largely a theoretical concept, and their experimental realization is demanding because they typically emerge at transitions between gapped phases carrying distinct higher topological invariants, which generally require engineered long-range hopping that is difficult to achieve in solid-state materials, particularly in higher dimensions~\cite{zhou2025topological}.

On a different front, photonic synthetic frequency dimensions~\cite{yu2025comprehensive,Yuan2018,OzawaPrice2019,Dutt2019,dutt2022creating,Yuan2021Review,yuan2016photonic,daweiwang2022Light,wang2025versatile,zeng2026hybrid,li2023direct,PhysRevLett.132.183802,leefmans2022topological}, realized through dynamically modulated ring resonators, provide a particularly promising platform for overcoming these challenges. In such systems, long-range couplings can be independently engineered via discrete radio-frequency modulations, enabling direct access to the long-range hopping processes required for critical topology, while the architecture naturally extends to higher synthetic dimensions. This raises a timely and intriguing question: Can one systematically develop an experimentally accessible protocol for realizing critical topological states across different photonic synthetic dimensions?

In this work, we fill this gap by proposing a photonic realization of critical topological states in synthetic frequency dimensions. Using equal-amplitude, opposite-phase trichromatic modulation in two coupled ring resonators, we demonstrate the emergence of critical topological photonic states, in which robust edge modes surprisingly persist within a gapless bulk continuum. In contrast to previously recognized topological photonics, whose nontrivial topology is characterized by quantized topological invariants, the critical topological states studied here possess ill-defined quantized topological invariants and can instead be diagnosed through the entanglement spectrum of the bulk wave functions, thus establishing an \textit{unconventional} bulk–boundary correspondence~\cite{PhysRevLett.101.010504,PhysRevLett.113.106801,PhysRevLett.113.060501,PhysRevB.84.205136,guo2025generalized}. We further investigate phase transitions between distinct critical topological photonic states, revealing that \textit{topology can reorganize criticality and drive phase transitions within critical regimes.} Finally, we extend these critical topological photonic states to 2D synthetic lattices and provide concrete electro-optic implementation schemes for both 1D and 2D settings, with experimentally realistic parameters within reach of current synthetic-frequency photonic platforms.

\begin{figure}[t]
  \centering
  \includegraphics[width=1.0\linewidth]{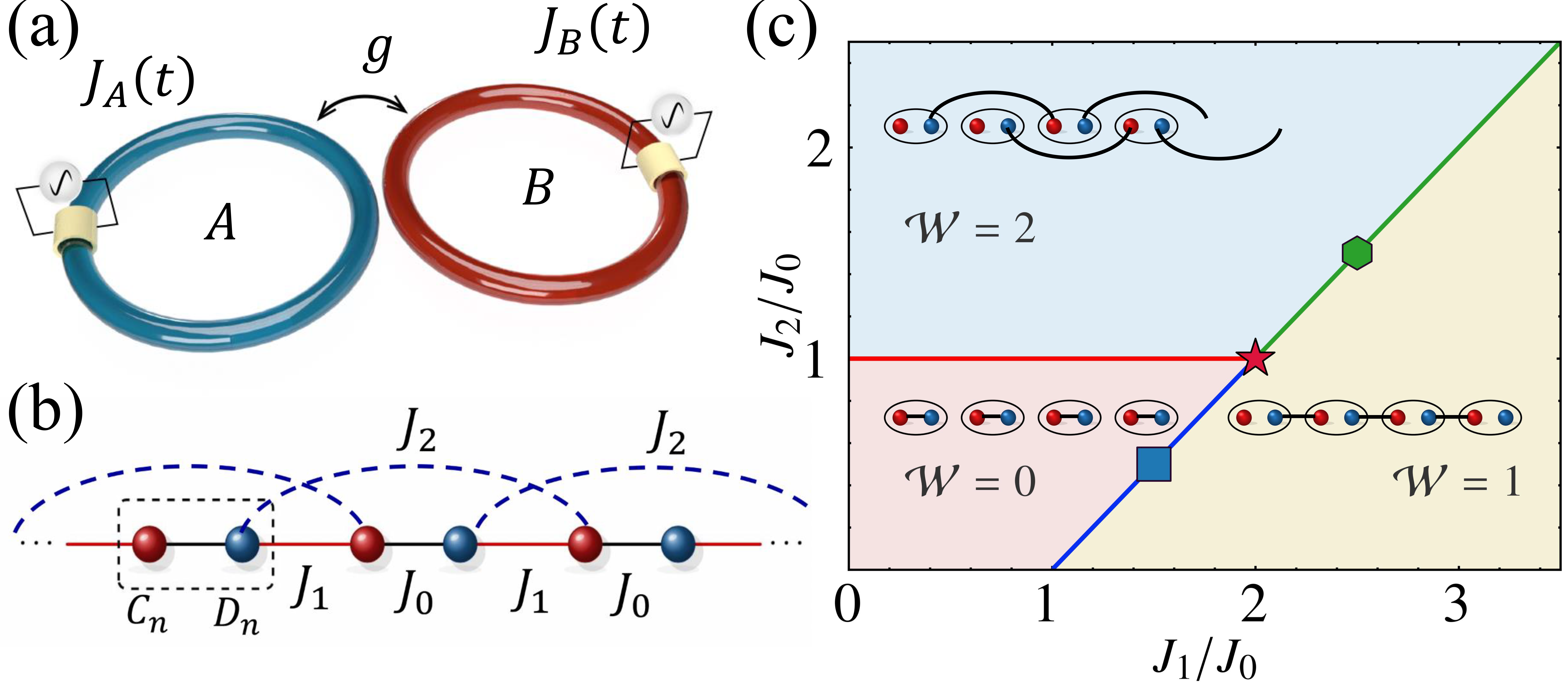}
\caption{
(a) Schematic of two coupled ring resonators, labeled A and B, with coupling strength $g$. Each ring contains an electro-optic modulator (EOM), shown in yellow. (b) Synthetic-frequency lattice in the supermode basis $\{C_n,D_n\}$ with intra-cell hopping $J_0$ and inter-cell hoppings $J_1$, $J_2$. (c) Topological phase diagram in the $(J_1/J_0, J_2/J_0)$ plane, exhibiting three gapped phases distinguished by the winding number $\mathcal{W}=0,1,2$, separated by three critical lines (blue, red, and green) that meet at a multicritical point (red star).
}
\label{figure1}
\end{figure}

\emph{System and model.}---We consider two identical ring resonators labeled A (blue) and B (red) in Fig.~\ref{figure1}(a)~\cite{peng2014parity,peng2014loss}. Each ring has length $L$ and group velocity $v_g$. In the absence of group-velocity dispersion, ring A (B) supports equally-spaced resonant modes defined as $A_n$ ($B_n$) at frequency $\omega_n = \omega_0 + n\Omega$, where $\omega_0$ is the reference frequency, $n$ is the mode index, and $\Omega = 2\pi v_g / L$ is the free spectral range (FSR) of the ring~\cite{li2023direct}.
Modes $A_n$ and $B_n$ with the same mode index $n$ can be coupled either by evanescent coupling or by a fiber coupler between the two rings, with coupling strength $g$.
One electro-optic modulator (EOM) is placed in each ring, and driven by equal-amplitude, opposite-phase trichromatic signals,
$J_A(t)=2\sum_{j=0}^{2}J_j\cos(\Omega_jt+\phi_j)$ and
$J_B(t)=-J_A(t)$, respectively~\cite{dutt2020higher,sridhar2025measuring}. Here, $\{J_j\}$, $\{\Omega_j\}$, and $\{\phi_j\}$ detote the modulation amplitudes, frequencies, and phases of the three RF components, respectively.
The system Hamiltonian reads
\begin{equation}
  \begin{aligned}
      H_0 =& \sum_n \omega_n (a^\dagger_{n} a_{n} + b^{\dagger}_{n} b_{n}) + g\sum_{n}(a^{\dagger}_{n} b_{n} + a_{n} b^{\dagger}_{n}) \\
    &+ \sum_{n,n'} \left(J_A(t)\, a^{\dagger}_{n}a_{n'} +
J_B(t)b_n^\dagger b_{n'}\right)\;.
  \end{aligned}
\end{equation}
Here, $a_{n}$ and $b_{n}$ ($a^{\dagger}_{n}$ and $b^{\dagger}_{n}$) are the annihilation (creation) operators for the cavity modes $A_n$ and $B_n$, respectively.
The coupling strength $g$ hybridizes the resonant modes at frequency $\omega_n$ into the symmetric supermode $C_n$ with frequency $(\omega_n + g)$ and the antisymmetric supermode $D_n$ with frequency $(\omega_n - g)$. These supermodes form a frequency lattice with alternating spacings $\Omega_0\equiv 2g$ and $\Omega_1\equiv\Omega-2g$.
We introduce operators $c_n = (a_n + b_n)/\sqrt{2}$ and $d_n = (a_n - b_n)/\sqrt{2}$ as the annihilation operators for the supermodes $C_n$ and $D_n$. We choose the three drive frequencies as $\Omega_0 = 2g$, $\Omega_1 = \Omega - 2g$, and $\Omega_2 = 2\Omega - 2g$ to match the inter-supermode spacings. Moving to the rotating frame through $\tilde c_n=c_ne^{i(\omega_n+g)t}$ and $\tilde d_n=d_ne^{i(\omega_n-g)t}$ removes the supermode onsite frequencies. Under the rotating-wave approximation, the remaining counter-rotating and off-resonant inter-supermode processes are neglected, yielding
\begin{equation}\label{eq:Hamiltonian}
      H = \sum_{n} \left( J_0\, \tilde{c}^{\dagger}_{n} \tilde{d}_{n} + J_1\, \tilde{d}^{\dagger}_{n} \tilde{c}_{n-1} + J_2\, \tilde{d}^{\dagger}_{n} \tilde{c}_{n-2}\right) + \mathrm{H.c.}
\end{equation}
Here, the RF phases have been chosen as $\phi_j=0$. This realizes a bipartite tight-binding chain with next-nearest-neighbor inter-sublattice hopping in the synthetic frequency dimension, as depicted in Fig.~\ref{figure1}(b). The hopping amplitudes $J_{1,2}$ are independently controlled by the two RF drive amplitudes with frequencies $\Omega_1, \Omega_2$, and $J_0$ is set as the energy unit. Fourier transforming Eq.~\eqref{eq:Hamiltonian} along the synthetic frequency axis yields the Bloch Hamiltonian $H(k_f) = h_x(k_f)\, \sigma_x + h_y(k_f)\, \sigma_y \;,$
where $k_f$ denotes the Bloch momentum in the synthetic frequency dimension (throughout this work, the subscript $f$ labels quantities in this synthetic space), $h_x(k_f) = J_0 + J_1 \cos(k_f) + J_2\cos(2k_f)$, and $h_y(k_f) = J_1 \sin(k_f) + J_2 \sin(2k_f)$. The chiral symmetry $\sigma_z H(k_f) \sigma_z = -H(k_f)$ places the model in the AIII class of the tenfold classification~\cite{ChiuTeo2016,ryu2010topological}. The boundary zero modes discussed below are protected as long as
perturbations preserve this chiral structure, i.e.,
$\{\delta H,\sigma_z\}=0$. Equivalently, the off-diagonal element $G(k_f) = h_x + ih_y = J_0 + J_1 e^{ik_f} + J_2 e^{2ik_f}$ defines an integer winding number $\mathcal{W} = \frac{1}{2\pi i}\oint dk_f\, \partial_{k_f}\ln G$. The resulting phase diagram, shown in Fig.~\ref{figure1}(c) contains three gapped phases $\mathcal{W}=0,1,2$, separated by critical lines along which $G(k_f)$ vanishes: $\mathcal{W}=0\leftrightarrow1$ (blue), $\mathcal{W}=0 \leftrightarrow 2$ (red), and $\mathcal{W}=1\leftrightarrow 2$ (green).

\begin{figure}
  \centering
  \includegraphics[width=1.0\linewidth]{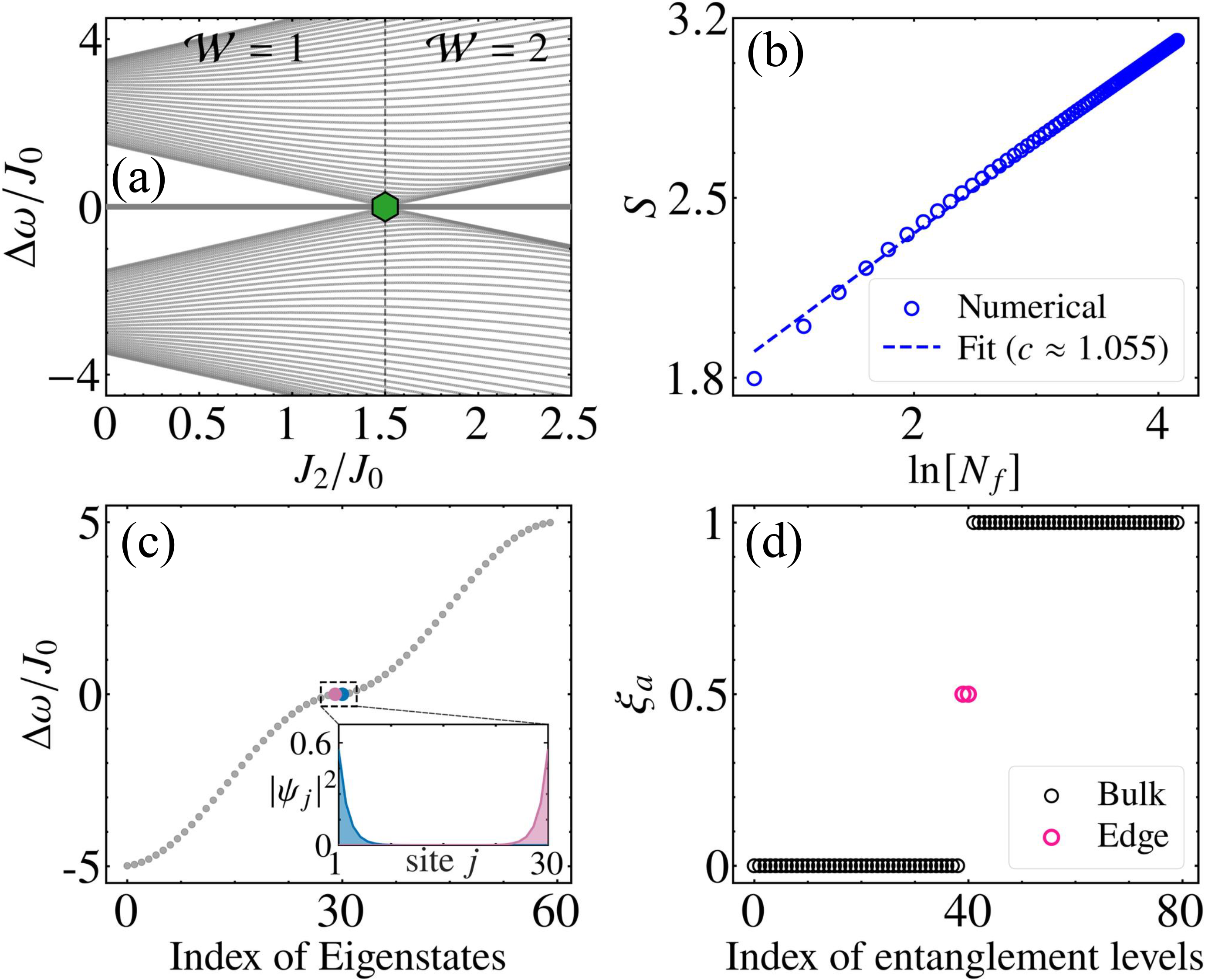} 
\caption{
Evidence for a critical topological photonic state on the 
$\mathcal W=1\leftrightarrow2$ transition line.
(a) Open-boundary spectrum as a function of $J_2/J_0$ for an extended
SSH chain with $N_f=40$ unit cells and $J_1/J_0=2.5$. The green hexagon marks the critical point $J_2/J_0=1.5$.
(b) Bipartite entanglement entropy scaling at the same critical point, computed under anti--periodic boundary conditions up to $N_f=2000$ unit cells. 
(c) Open-boundary energy spectrum at the critical point for $N_f=30$ unit cells. The inset shows the corresponding wave-function intensities $|\psi_j|^2$ along the synthetic-frequency chain.
(d) Single-particle entanglement spectrum with $N_f=80$ unit cells.
}
\label{figure2}
\end{figure}

\emph{Critical topological photonic state.}---The central theme of this work is to uncover topological phenomena emerging at criticality in synthetic frequency dimensions. To this end, we trace the evolution of the open-boundary eigenfrequency spectrum by fixing $J_1/J_0=2.5$ and tuning $J_2/J_0$ from $0$ to $2.5$. As shown in Fig.~\ref{figure2}(a), the zero-frequency mode continuously evolves from the $\mathcal{W}=1$ phase to the $\mathcal{W}=2$ phase, accompanied by a bulk-gap closing at $(J_1/J_0,J_2/J_0)=(2.5,1.5)$. While band-touching singularities frequently arise in topological photonics~\cite{huang2011dirac,Sakoda:12}, the closing of the bulk gap alone does not establish the emergence of criticality. To determine the nature of this band-touching point, we examine the scaling of the half-chain entanglement entropy $S$. For a 1D conformally invariant critical point, conformal field theory predicts the universal scaling relation $S(N_f)=\frac{c}{3}\ln N_f+s_0$, where $c$ denotes the central charge and $s_0$ is a nonuniversal constant~\cite{calabrese2004entanglement}. As shown in Fig.~\ref{figure2}(b), the observed logarithmic scaling firmly establishes that the $\mathcal{W}=1\leftrightarrow2$ transition realizes a genuine conformal critical point rather than an accidental band touching.

Having established the critical nature of the transition, we next examine whether topological boundary modes persist in the critical regime. Figure~\ref{figure2}(c) shows the eigenfrequency spectrum of an open synthetic-frequency chain with $N_f=30$ unit cells at the critical band-touching point. Remarkably, two degenerate states remain pinned at zero frequency despite being embedded within the gapless bulk continuum. The inset further reveals that their wave functions are strongly localized at the system boundaries, indicating the presence of twofold-degenerate boundary modes. To assess their physical origin, we note that in conventional topological photonics, boundary states are protected by a quantized bulk topological invariant via the bulk–boundary correspondence~\cite{Ozawa2019,HasanKane2010}. At criticality, however, the bulk invariant becomes ill-defined since the band gap closes. In this context, the notion of topology can instead be diagnosed through the entanglement spectrum of the bulk state, following the recently proposed generalized Li–Haldane correspondence~\cite{guo2025generalized} (see End Matter for details). Within this framework, entanglement eigenvalues pinned at $\xi=1/2$ indicate the presence of topologically protected boundary modes. As shown in Fig.~\ref{figure2}(d), the $\mathcal{W}=1\leftrightarrow2$ critical point exhibits a twofold degeneracy at $\xi=1/2$ in the bulk entanglement spectrum, in one-to-one correspondence with the boundary-mode degeneracy observed in Fig.~\ref{figure2}(c). These results demonstrate that the boundary modes embedded in the gapless continuum are topologically protected, rather than arising from accidental localization. Because the entanglement spectrum is determined entirely by the single-particle band projector, this diagnostic admits quantum–classical correspondence and can be reconstructed from phase-sensitive measurements on the synthetic-frequency platform (see End Matter and SM Sec.~II~\cite{SM} for the measurement protocol). The robustness of these critical boundary modes against symmetry-preserving disorder is further discussed in the SM Sec.~IV~\cite{SM}.

We now turn to the \textit{physical picture} underlying critical topology. In
conventional topological photonics, boundary modes arise from mass inversion: a
domain wall in the mass term binds a localized state~\cite{PhysRevD.13.3398}.
This picture fails once the gap closes and the mass vanishes. Writing
$k_f=\pi+q$ and expanding the off-diagonal element gives
$G\simeq m+i\kappa q-Aq^{2}$, with $m=J_0-J_1+J_2$, $\kappa=2J_2-J_1$, and
$A=2J_2-J_1/2$. The critical lines are fixed by $m=0$, along which
$\kappa=J_2-J_0$ changes sign at the multicritical point while $A>0$ stays
finite. The zero-mode equation $(A\partial_x^{2}+\kappa\partial_x)\psi=0$ then
yields $\psi(x)\sim \exp(-\kappa x/A)$, bound at a termination only for
$\kappa>0$, i.e. $J_2>J_0$. Criticality alone is thus insufficient: the sign of
the kinetic coefficient, not a mass gap, selects the
$\mathcal{W}=1\leftrightarrow2$ branch as topological and sets the localization
length $A/|\kappa|$, which diverges as $\kappa \to 0$.
This \emph{kinetic-inversion} mechanism~\cite{verresen2020} (see also SM
Sec.~I~\cite{SM}) is the gapless counterpart of Jackiw--Rebbi.

\begin{figure}
  \centering
  \includegraphics[width=1.0\linewidth]{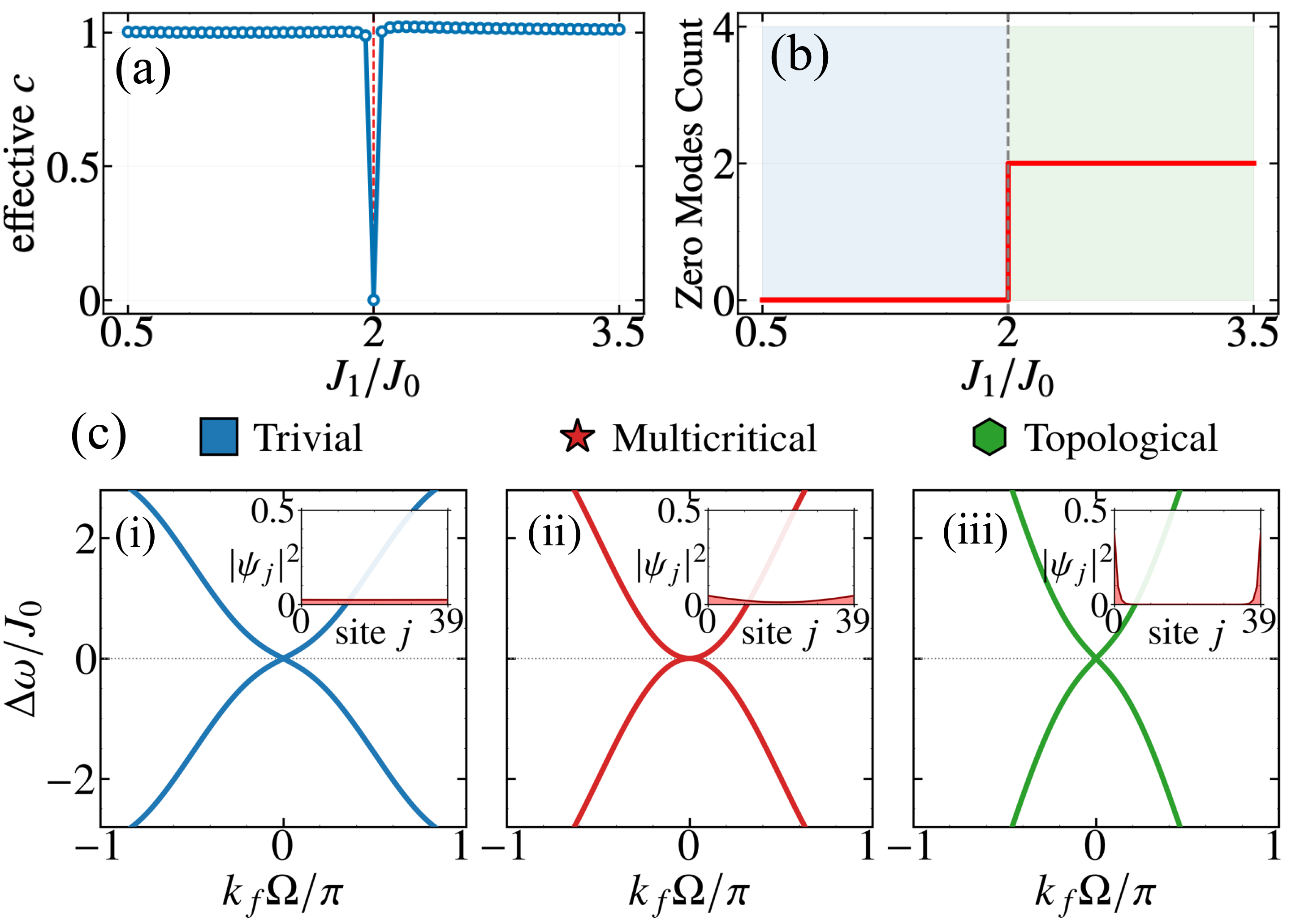}
\caption{(a) Effective central charge scaling along the critical trajectory $J_2=J_1-J_0$. The system size is $N_f=2000$ unit cells. (b) Number of boundary zero modes as a function of $J_1/J_0$ along the critical line. (c, i)--(c, iii) Schematic photonic single-particle energy spectra: linear dispersion in the topologically trivial regime ($J_1/J_0=1.5$, $J_2/J_0=0.5$), quadratic band touching 
at the multicritical point ($J_1/J_0=2.0$, $J_2/J_0=1.0$), and linear dispersion at shifted momentum in the topologically nontrivial regime ($J_1/J_0=3.0$, $J_2/J_0=2.0$).}
\label{figure3}
\end{figure}  

\emph{Topologically enforced multicriticality in photonics.}---After identifying critical topological states in photonic systems, we next examine how nontrivial topology is reorganized along critical phase boundaries, namely at multicritical points separating topologically distinct critical states. This phenomenon is fundamentally different from that in conventional topological photonics, where multicritical points typically arise as intersections of distinct universality classes characterized by different central charges.

Specifically, we consider the trajectory $J_2/J_0=J_1/J_0-1$ that follows the critical lines, and tune $J_1/J_0$ from 1.5 to 2.5 in Fig.~\ref{figure1}(c). The central charge extracted from the entanglement entropy scaling is shown in Fig.~\ref{figure3}(a). Notably, the central charge remains $c = 1$ throughout but drops abruptly to zero at $J_1/J_0=2.0$, signaling a transition between distinct critical lines that nonetheless share the same central charge. To further characterize this transition, we analyze the number of topological edge modes via the bulk entanglement spectrum, and find that the number of topological zero modes also changes discontinuously at $J_1/J_0=2.0$: the adjacent critical regimes host zero and two boundary modes, respectively. This topologically enforced transition can be further visualized by the evolution of the eigenfrequency spectrum along the critical state, as shown in Fig.~\ref{figure3}(c). The band dispersion evolves from linear in the trivial critical state to quadratic at the multicritical point, and back to linear in the critical topological state, while boundary modes emerge only after crossing the multicritical point into the critical topological state. Therefore, this provides a new mechanism for multicriticality, driven by changes in the topology of neighboring critical lines, which lies beyond the conventional framework of topological photonics.

\emph{Extension to 2D synthetic frequency space.}---The synthetic-frequency construction can be naturally generalized to higher dimensions by combining a real-space resonator array with the frequency lattice associated with each ring resonator. As a concrete example, we consider a hybrid 2D lattice described by an extended Qi-Wu-Zhang (QWZ) model with long-range hopping~\cite{PhysRevB.74.085308,verresen2020,PhysRevB.102.134213},
\begin{equation}
\begin{aligned}
\mathcal{H}(k_x,k_f)
=&\left(\sin k_f+\lambda\sin2k_f\right)\sigma_x
+\sin k_x\,\sigma_y  \\
&+\left(M-\cos k_f-\lambda\cos2k_f-\cos k_x\right)\sigma_z .
\end{aligned}
\label{eq:2D_extended_QWZ}
\end{equation}
Here, $k_x$ denotes the momentum along the real-space direction, while $k_f$ labels the momentum along the synthetic frequency dimension. The Pauli matrices act on the two resonator modes $(a_{i,n},b_{i,n})$. Nearest-neighbor coupling along the real-space direction is provided by the resonator array, whereas the nearest- and next-nearest-neighbor couplings along the frequency dimension are generated by the corresponding electro-optic modulation tones. Details of the real-space hopping matrices, the global phase diagram, and potential experimental implementations are provided in Sec.~III of the SM~\cite{SM}.

\begin{figure}[!t]
  \centering
  \includegraphics[width=1.0\linewidth]{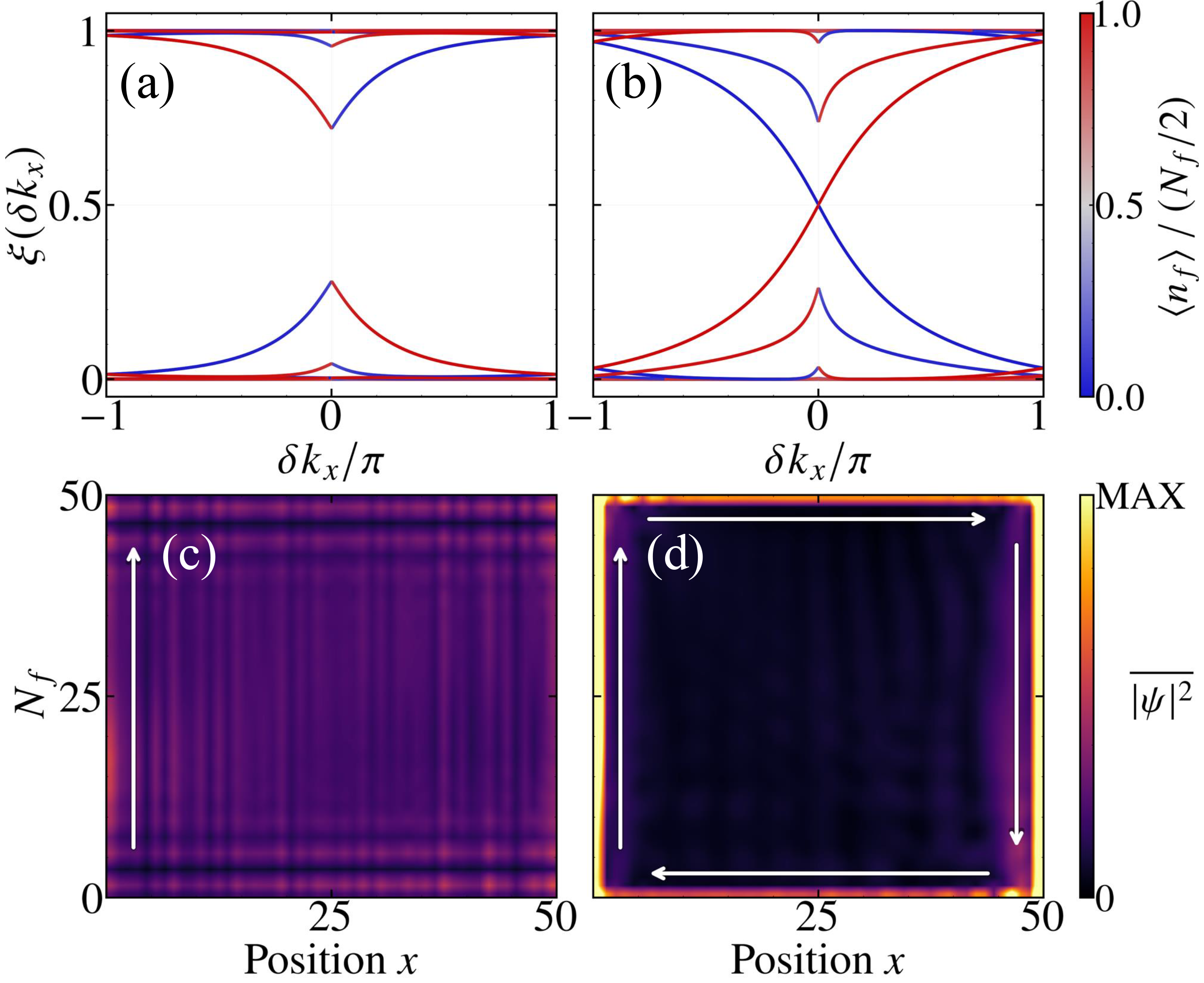}
\caption{2D extension of critical topological photonics.
(a, b)Single-particle entanglement spectra of the extended QWZ model
at $\lambda=1$ for (a) the conventional critical point $M=3$ and
(b) the critical topological point $M=-1$.  The finite frequency
direction contains $N_f=40$ unit cells, the momentum $k_x$ is sampled with
$N_{k_x}=800$, and the horizontal axis is measured relative to the
gap-closing momentum, $\delta k_x=k_x-k_x^\ast$, with
$k_x^\ast=0$ for $M=3$ and $k_x^\ast=\pi$ for $M=-1$.
The color indicates the mean frequency position of the entanglement
eigenstate.  (c, d) Time-averaged probability density
$\overline{\rho}(x,y)$ of the same edge-localized Gaussian wavepacket
evolved in a $50\times50$ open lattice.}
\label{figure4}
\end{figure}

This higher-dimensional setting enables the realization of critical states with distinct topological characteristics within a single synthetic platform. In Fig.~\ref{figure4}, we focus on two representative points along the critical line. The point $(M, \lambda)=(3,1)$ corresponds to a topologically trivial critical state separating the $C=0$ and $C=-1$ Chern insulating phases, whereas $(M, \lambda)=(-1,1)$ realizes a critical topological state between the $C=1$ and $C=2$ Chern insulators. The distinction between these two critical states is directly encoded in their entanglement spectra, computed under periodic boundary conditions along the real-space direction and open boundary conditions along the synthetic frequency dimension. As shown in Figs.~\ref{figure4}(a, b), the trivial critical state exhibits no chiral boundary modes, whereas the topological critical state hosts a pair of chiral entanglement modes crossing at $\xi=1/2$. Moreover, this topological distinction is further reflected in the real-space dynamics. As illustrated in Figs.~\ref{figure4}(c, d), an edge-localized Gaussian wavepacket rapidly disperses into the gapless bulk for $(M, \lambda)=(3,1)$, whereas it remains confined to the boundary for $(M, \lambda)=(-1,1)$. Therefore, the proposed synthetic-frequency architecture offers a versatile platform for extending critical topological photonics to higher-dimensional synthetic lattices.

\emph{Discussion and Concluding remarks.}---Beyond advancing the conceptual understanding of topological photonics, our results also point toward novel critical topological photonic devices and suggest a possible route to critical topological sensing distinct from conventional topological sensors~\cite{PhysRevLett.129.090503,xuepeng2026prl}.  Rather than approaching a gap closing where ordinary gap protection deteriorates, the present setting places the sensor at a critical point that still carries a topological boundary structure.  This may allow enhanced critical response to coexist with symmetry-protected noise filtering: symmetry-preserving perturbations are suppressed, while signals that break or shift the protected structure, such as local frequency shifts, can be selectively detected. Further exploration of these potential applications is left for future work. 

In summary, we have proposed an experimentally accessible route to critical topological photonics in synthetic frequency dimensions. Building on two coupled ring resonators with trichromatic electro-optic modulation, we study an extended SSH chain with independently tunable long-range hopping, providing a minimal platform for realizing critical topology in photonics. Unlike conventional topological photonics, the critical topological photonic states identified here feature robust topological boundary modes that coexist with a gapless bulk continuum. The diagnosis of such critical topology is also fundamentally unconventional: ordinary quantized topological invariants become ill-defined, and topological boundary modes are instead faithfully encoded in degenerate midgap states of the bulk entanglement spectrum. We further investigate topologically enforced multicriticality between distinct critical topological photonic states, where topology actively drives a phase transition within the critical regime. Guided by the same design principle, the critical topological physics revealed above naturally extends to 2D synthetic lattices. Additionally, we have proposed concrete electro-optic implementations with experimentally realistic parameters for both the one- and 2D settings, indicating that the proposed critical topological photonic states are within reach of existing synthetic-frequency platforms.

\textit{Acknowledgement}:
We thank Professors~Luqi Yuan, Xiaoze Liu, Meng Xiao for helpful discussions.
X.-J. Yu was supported by the National Natural Science Foundation of China (Grant No.12405034) and a start-up grant from Eastern Institute of Technology, Ningbo. F. N. is supported in part by the Japan Science and Technology Agency (JST) [via the CREST Quantum Frontiers program Grant No. JPMJCR24I2, the Quantum Leap Flagship Program (Q-LEAP), and the Moonshot R\&D Grant No. JPMJMS2061].

\let\oldaddcontentsline\addcontentsline
\renewcommand{\addcontentsline}[3]{}
\bibliography{main.bib}

\appendix
\section{\large{End Matter}}
\twocolumngrid

\begin{figure}[!h]
  \centering
  \includegraphics[width=0.75\linewidth]{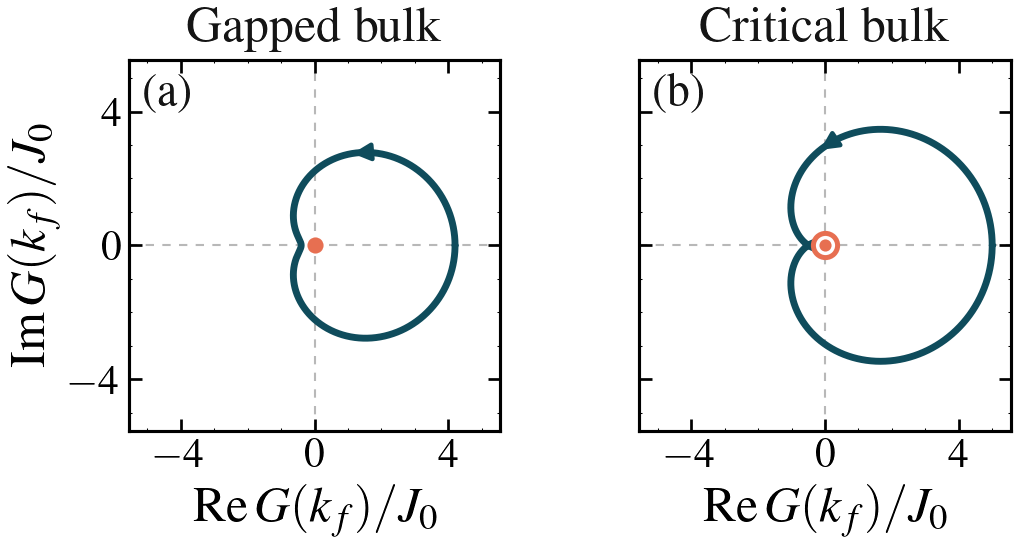}
  \caption{Complex-plane diagnosis of the winding-number breakdown. (a) At the gapped point $(J_1/J_0,J_2/J_0)=(2.3,0.9)$, $G(k_f)$ avoids the origin and the winding number is well defined. (b) At the critical point $(J_1/J_0,J_2/J_0)=(2.5,1.5)$, $G(k_f^\ast)=0$, so $\arg G(k_f)$ is singular and the conventional winding number is ill defined.}
  \label{figure5}
\end{figure}

\emph{Differences between topological photonics and critical topological photonics.}---Conventional topological photonics is rooted in the bulk-boundary correspondence: a nontrivial topological invariant defined for the gapped bulk bands predicts the existence of topologically protected boundary modes at an interface with a topologically distinct medium. Critical topological photonics differs in a fundamental way. At a critical point, the bulk gap closes, so the usual topological invariant is no longer well defined. This can be seen directly from $G(k_f)$ in Eq.~\eqref{eq:Hamiltonian}, where $G(k_f)\equiv h_x(k_f)+i h_y(k_f)
=J_0+J_1e^{ik_f}+J_2e^{2ik_f}$. In the gapped regions, the trajectory of $G(k_f)$ avoids the origin, and the winding number counts how many times this loop encircles the origin. At criticality, however, the loop touches the origin, i.e., $G(k_f^\ast)=0$, which corresponds to a bulk band closing and invalidates the conventional winding-number definition, see Fig.~\ref{figure5}. The schematic winding loops, therefore, illustrate why a critical point cannot be characterized simply by assigning the winding number of either adjacent gapped phase.

However, the failure of the conventional invariant does not imply the absence of bulk-edge correspondence in critical topological photonics. Along the critical line separating the $\mathcal W=1$ and $\mathcal W=2$ phases, chiral symmetry pins zero-frequency boundary modes even though the bulk spectrum is gapless. Thus, the relevant correspondence is no longer the usual ``gapped bulk invariant--topologically protected edge state'' relation. Instead, the relevant information is transferred to the bulk entanglement structure, where the degeneracy of the boundary zero modes is faithfully encoded in the entanglement spectrum via the recently proposed generalized Li--Haldane correspondence~\cite{PhysRevLett.113.106801,guo2025generalized}. In this framework, the characteristic midgap structure of the entanglement spectrum provides a robust diagnostic of critical topology. For a half-filled periodic chain, we
construct the correlation matrix from the occupied single-particle states,
\begin{equation}
C_{ij}=\sum_{m\in {\rm occ}} \psi_m(i)\psi_m^*(j).
\label{eq:app_C_full}
\end{equation}
Restricting $C$ to a spatial subsystem $A$ gives the reduced correlation
matrix $C^A$.  Its eigenvalues define the single-particle entanglement
spectrum,
\begin{equation}
C^A|\varphi_\alpha\rangle
=
\xi_\alpha|\varphi_\alpha\rangle,
\qquad
0\leq \xi_\alpha\leq1 .
\label{eq:app_ES_definition}
\end{equation}
Equivalently, the corresponding entanglement energies are $\varepsilon_\alpha = \ln\frac{1-\xi_\alpha}{\xi_\alpha}$,
so that $\xi_\alpha=1/2$ corresponds to zero entanglement energy.
The key point is that the entanglement cut creates virtual boundaries in an otherwise periodic system.  Therefore, midgap levels pinned near
$\xi=1/2$ diagnose boundary degrees of freedom encoded in the bulk wave functions, without relying on a physical edge termination. In this sense, \textit{the entanglement spectrum restores a bulk-boundary correspondence at criticality and distinguishes a topologically nontrivial critical point from an ordinary band closing.}

\emph{Experimental implementation.}---The extended 1D SSH synthetic lattice can be implemented using two coupled fiber-ring resonators, each containing an EOM.  For a ring length $L=10~{\rm m}$ and group index $n_g\simeq1.5$, the free spectral range is $\Omega/(2\pi)\simeq20~{\rm MHz}$. An inter-ring coupling $g/(2\pi)=4~{\rm MHz}$ yields the resonant modulation tones $\Omega_{0,1,2}/(2\pi)=(8,12,32)~{\rm MHz}$, corresponding to $J_0,J_1,J_2$. The critical point shown in Fig.~\ref{figure2} of the main text is obtained with $(J_0,J_1,J_2)/(2\pi)=(0.20,0.50,0.30)~{\rm MHz}$, requiring EOM modulation indices $\beta=(0.02,0.05,0.03)$, within standard lithium-niobate EOM capabilities.

\let\addcontentsline\oldaddcontentsline
\onecolumngrid

\clearpage
\newpage

\widetext

\begin{center}
\textbf{\large Supplemental Material for ``Critical Topological Photonics in Synthetic Dimensions''}
\end{center}

\maketitle

\renewcommand{\thefigure}{S\arabic{figure}}
\setcounter{figure}{0}
\renewcommand{\theequation}{S\arabic{equation}}
\setcounter{equation}{0}
\renewcommand{\thesection}{\Roman{section}}
\renewcommand{\thesubsection}{\Alph{subsection}}
\setcounter{section}{0}
\setcounter{secnumdepth}{4}

\makeatletter
\@removefromreset{equation}{section}
\makeatother

\addtocontents{toc}{\protect\setcounter{tocdepth}{0}}
{
\tableofcontents
}

\section{Physical origin of critical topological photonics: kinetic inversion}
\label{sec:SM_kinetic_inversion}
In this section, we provide a simple physical interpretation of the 
critical topological photonics discussed in the main text.  The boundary modes identified at the critical topological points in the
main text do not originate from the conventional mass-inversion (Jackiw--Rebbi) mechanism of gapped topological
insulators~\cite{PhysRevD.13.3398}. Instead, they arise from a \emph{kinetic inversion} near the critical
point~\cite{verresen2020}. 

We first recall the mass-inversion mechanism in a gapped topological insulator. Near a transition between a trivial phase and a topological phase, the low-energy theory is commonly described by a Dirac Hamiltonian with a spatially varying mass $m(x)$. For a spatial interface where the mass changes sign, $m(x\!\to\!-\infty)\,m(x\!\to\!+\infty)<0$, the resulting domain wall binds a zero-energy mode. The localization length of this mode is set by the inverse bulk gap, namely by the inverse magnitude of the asymptotic mass. This is the standard mass-inversion, or band-inversion, mechanism underlying the bulk--boundary correspondence in gapped topological phases.
Note that in Ref.~\cite{bliokh2019topological} the mass is replaced by the helicity from Maxwell equations.

The critical topological state considered here is fundamentally different. At a critical point the bulk gap vanishes, so the mass-inversion picture is no longer applicable. To expose the mechanism that replaces it, we expand the off-diagonal Bloch element
$G(k_f)=J_0+J_1 e^{ik_f}+J_2 e^{2ik_f}$ in Eq.~(2) of the main text around the gap-closing momentum $k_f=\pi$. Writing $k_f=\pi+q$ and expanding to leading orders gives
\begin{equation}
G(\pi+q)
\simeq
m+i\kappa\, q-A\,q^2 ,
\label{eq:SM_continuum_G}
\end{equation}
with $m=J_0-J_1+J_2$, $\kappa=-J_1+2J_2$, and $A=2J_2-J_1/2$.
Here $m$ controls the gap opening and $\kappa$ is the linear kinetic coefficient. 

The condition \(m=0\), equivalently \(J_1=J_0+J_2\), defines the diagonal gap-closing trajectory. For \(J_2<J_0\) it is the \(W=0\leftrightarrow1\) branch, whereas for \(J_2>J_0\) it is the $W=1\leftrightarrow2$ branch; the two meet at \(J_2=J_0\). Along the line $W=1\leftrightarrow2$ the kinetic coefficient $\kappa=-J_1+2J_2$ remains generically nonzero. It is this nonvanishing $\kappa$, rather than a mass gap, that governs the boundary physics of the critical topological state. 

The multicritical point $J_1=2J_0$, $J_2=J_0$ corresponds to the special case in which $\kappa$ also vanishes; here the leading dispersion reduces to a quadratic band touching, and distinct critical branches are reorganized.
\begin{figure*}
\centering
\begin{tikzpicture}[
    >=Latex,
    line cap=round,
    line join=round,
    font=\sffamily,
    x=1cm,
    y=1cm
]

\definecolor{diskgray}{RGB}{235,235,235}
\definecolor{leftbg}{RGB}{226,246,243}
\definecolor{rightbg}{RGB}{236,232,250}
\definecolor{teal}{RGB}{0,135,135}
\definecolor{purple}{RGB}{82,65,165}
\definecolor{redwave}{RGB}{220,35,45}
\definecolor{blackcurve}{RGB}{20,20,20}
\definecolor{lightaxis}{RGB}{150,150,150}

\tikzset{
    boxframe/.style={gray!55,thin},
    zeroline/.style={lightaxis,thin},
    intline/.style={gray!70,dashed,thick},
    profile/.style={blackcurve,very thick,smooth,samples=200},
    kprofile/.style={blackcurve,very thick,dashed,smooth,samples=200},
    wavefill/.style={redwave,opacity=0.62,smooth,samples=200},
    waveline/.style={redwave,very thick,smooth,samples=200},
    panelaxis/.style={->,thick,blackcurve}
}

\node[font=\bfseries\large] at (-4.10,2.55) {(a)};
\node[font=\bfseries\large] at ( 5.10,2.55) {(b)};

\begin{scope}[shift={(0,0)}]

    \node[font=\large] at (0,2.30) {mass inversion};

    \fill[leftbg]  (-3.75,-1.85) rectangle (0,2.00);
    \fill[rightbg] (0,-1.85) rectangle (3.75,2.00);
    \draw[boxframe] (-3.75,-1.85) rectangle (3.75,2.00);

    \draw[zeroline] (-3.75,0) -- (3.75,0);
    \draw[intline] (0,-1.85) -- (0,2.00);
    \draw[panelaxis] (-3.88,-1.85) -- (3.93,-1.85)
        node[right,font=\Large] {$x$};
    \draw[panelaxis] (-3.75,-1.95) -- (-3.75,2.20);
    \node[font=\normalsize,anchor=east,xshift=-2pt]
    at (-3.75,0) {$0$};
    \node[teal,font=\normalsize] at (-1.88,1.80) {Vacuum};
    \node[purple,font=\normalsize] at (1.88,1.80) {Gapped topology};

    \fill[wavefill,domain=-3.00:3.00]
        plot(\x,{1.05*exp(-1.09*(\x)^2)}) --
        (3.00,0) -- (-3.00,0) -- cycle;
    \draw[waveline,domain=-3.00:3.00]
        plot(\x,{1.05*exp(-1.09*(\x)^2)});
    \node[redwave,font=\large,anchor=west] at (-3.60,0.85) {$|\psi(x)|^2$};

    \draw[profile,domain=-3.60:3.60]
        plot(\x,{1.55*tanh(1.03*\x)});
    \node[font=\large,anchor=west] at (1.80,0.90) {$m(x)$};

\end{scope}

\begin{scope}[shift={(9.2,0)}]

    \node[font=\large] at (0,2.30) {kinetic inversion};

    \fill[leftbg]  (-3.75,-1.85) rectangle (0,2.00);
    \fill[rightbg] (0,-1.85) rectangle (3.75,2.00);
    \draw[boxframe] (-3.75,-1.85) rectangle (3.75,2.00);

    \draw[zeroline] (-3.75,0) -- (3.75,0);
    \draw[intline] (0,-1.85) -- (0,2.00);
    \draw[panelaxis] (-3.88,-1.85) -- (3.93,-1.85)
        node[right,font=\Large] {$x$};
    \draw[panelaxis] (-3.75,-1.95) -- (-3.75,2.20);
    \node[font=\normalsize,anchor=east,xshift=-2pt]
    at (-3.75,0) {$0$};
    \node[teal,font=\normalsize] at (-1.88,1.80) {Vacuum};
    \node[purple,font=\normalsize] at (1.88,1.80) {Critical topology};

    \fill[wavefill,domain=-3.15:2.85]
        plot(\x,{1.00*exp(-0.84*(\x+0.27)^2)}) --
        (2.85,0) -- (-3.15,0) -- cycle;
    \draw[waveline,domain=-3.15:2.85]
        plot(\x,{1.00*exp(-0.84*(\x+0.27)^2)});
    \node[redwave,font=\large,anchor=west] at (-3.60,0.85) {$|\psi(x)|^2$};

    \draw[profile,domain=-3.63:3.63]
        plot(\x,{1.55*(tanh(1.03*\x)-1)/2});
    \node[font=\large,anchor=west] at (1.28,-0.55) {$m(x)$};

    \draw[kprofile,domain=-3.63:3.63]
        plot(\x,{1.55*tanh(1.03*\x)});
    \node[font=\large,anchor=west] at (1.65,1.20) {$\kappa(x)$};

\end{scope}
\end{tikzpicture}
\caption{
Physical origin of the critical topological photonic states from the kinetic inversion.
(a)~In a conventional gapped topological interface, a localized zero mode
is generated by a sign change of the mass term $m(x)$.
(b)~In a critical topological interface, the boundary mode is instead
generated by a sign change of the kinetic coefficient $\kappa(x)$ while
the mass remains tuned to the critical line $m\approx 0$.
This provides a gapless analogue of the Jackiw--Rebbi domain-wall
mechanism~\cite{verresen2020}.
}
\label{fig:kinetic_inversion}
\end{figure*}
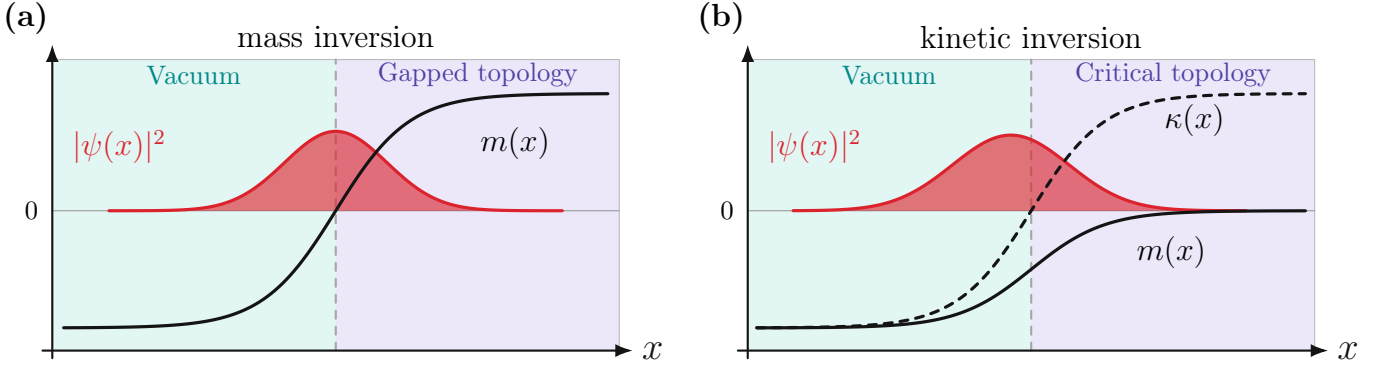
For a spatial interface the corresponding low-energy operator reads,
up to an overall sign convention,
\begin{equation}
\mathcal{D}
=
-A\,\partial_x^2-\kappa(x)\,\partial_x+m(x).
\label{eq:SM_kinetic_operator}
\end{equation}
In a gapped topological interface, the zero mode is produced by a sign
change of $m(x)$, as shown schematically in
Fig.~\ref{fig:kinetic_inversion}(a). Along a critical topological
interface, by contrast, the system sits on the critical line $m(x)=0$
while the kinetic coefficient $\kappa(x)$ interpolates between
opposite asymptotic signs, see Fig.~\ref{fig:kinetic_inversion}(b). The zero-mode equation then
reduces to
\begin{equation}
\bigl[-A\,\partial_x^2
      -\kappa(x)\,\partial_x\bigr]\psi(x)=0 ,
\label{eq:SM_zero_mode_kinetic}
\end{equation}
which, for a constant $\kappa(x)\equiv \kappa$ with $x\rightarrow \infty$, supports an exponentially localized solution
\begin{equation}
\psi(x)\sim
\exp\!\left[-\frac{\kappa}{A} x\right],
\label{eq:SM_kinetic_zero_mode}
\end{equation}
provided the sign of $\kappa/A$ is such that the wave function decays away from the interface on both sides. Thus the boundary localization is controlled by the kinetic coefficient $\kappa(x)$ rather than by a bulk mass gap.

This mechanism is the kinetic analogue of the Jackiw--Rebbi mass-domain-wall mechanism. In the conventional gapped case, the interface is topological because the mass term changes sign; in the critical case, it is topological because the kinetic coefficient
changes sign when passing through the critical point. The resulting boundary mode is therefore not a remnant of a finite-gap topological band structure, but an intrinsic boundary signature of the critical line. 

For the extended SSH chain, this picture explains why different critical lines in the phase diagram Fig.~1(c) in the main text are topologically inequivalent. The $\mathcal{W}=0\leftrightarrow1$ and $\mathcal{W}=0\leftrightarrow2$ critical lines border the trivial sector and do not retain protected boundary modes at criticality. The $\mathcal{W}=1\leftrightarrow2$ critical line, by contrast, is separated from these trivial branches by the critical point, across which $\kappa$ changes sign. This kinetic inversion provides the physical origin of the critical topological zero modes reported in the main text.

\section{Entanglement-spectrum diagnosis of critical topology}
\label{sec:entanglement_spectrum}

\subsection{Numerical construction of the entanglement spectrum}
\label{app:ES_calculation}

We now compute the single-particle entanglement spectrum from the
finite-size extended SSH Hamiltonian [Eq.~(2) of the main text],
\begin{equation}
H=
\sum_n
\left(
J_0 \tilde c_n^\dagger \tilde d_n
+
J_1 \tilde d_n^\dagger \tilde c_{n-1}
+
J_2 \tilde d_n^\dagger \tilde c_{n-2}
\right)
+{\rm h.c.},
\label{eq:app_real_space_SSH}
\end{equation}
on a chain of $N_f$ unit cells with basis
$\Psi=(\tilde c_1,\tilde d_1,\ldots,
\tilde c_{N_f},\tilde d_{N_f})^T$. Anti-periodic boundary conditions
(APBC) are imposed so that the discrete momentum grid avoids the isolated gap-closing point at criticality. The Hamiltonian is diagonalized to yield $2N_f$ eigenstates $|\psi_\alpha\rangle$ with eigenvalues ordered increasingly. At half filling the occupied subspace consists of the $N_f$ lowest-frequency states, and the equal-time correlation matrix is the projector
\begin{equation}
C_{ij}
=
\sum_{\alpha=1}^{N_f}
\psi_{\alpha,i}\,\psi_{\alpha,j}^{*},
\label{eq:app_C_full}
\end{equation}
where $i,j$ run over both unit-cell and sublattice indices.
The subsystem $A$ is chosen as the first $N_A=N_f/2$ unit cells (including both sublattices). The restricted correlation matrix $C^A=C\big|_{i,j\in A}$ has dimension $2N_A\times 2N_A$, and its eigenvalues define the single-particle entanglement spectrum,
\begin{equation}
C^A|\varphi_\alpha\rangle
=
\xi_\alpha|\varphi_\alpha\rangle ,
\qquad
0\leq \xi_\alpha\leq1 .
\label{eq:app_CA_diag}
\end{equation}
The corresponding entanglement energies
$\varepsilon_\alpha=\ln[(1-\xi_\alpha)/\xi_\alpha]$
vanish at $\xi_\alpha=1/2$, so midgap entanglement levels signal zero modes of the entanglement Hamiltonian. For Fig.~2(d) of the main text we use the critical parameters $J_1/J_0=2.5$, $J_2/J_0=1.5$ with $N_f=80$ unit cells. The two eigenvalues pinned at $\xi=1/2$ are the virtual-cut counterparts of the two physical zero modes found under open boundary conditions, confirming that these midgap entanglement modes are encoded in the bulk critical wave functions.


\subsection{Quantum-classical correspondence of the entanglement spectrum}
\label{app:ES_photonic_interpretation}

The entanglement spectrum in Eqs.~\eqref{eq:app_C_full}--\eqref{eq:app_CA_diag}
is used here as a single-particle wave-function diagnostic. It should not be interpreted as a direct measurement of photon-photon many-body entanglement. For a quadratic fermionic band problem, the many-body ground-state entanglement spectrum is fully determined by the occupied-band projector, or equivalently by the single-particle correlation matrix $C=P_{\rm occ}$. Therefore, the required input is the set of band eigenvectors, not an interacting many-body state.

A linear photonic lattice governed by the same effective Hamiltonian has
the same single-particle eigenmodes. Hence, after reconstructing the
photonic band eigenvectors, one can construct the same projector
$P_{\rm occ}$, restrict it to a virtual subsystem $A$, and obtain
$\{\xi_\alpha\}$ by the procedure defined in
Eqs.~\eqref{eq:app_C_full}--\eqref{eq:app_CA_diag}. The resulting
entanglement spectrum is thus the spectrum of the auxiliary free-fermion
projector associated with the photonic band structure.

Accordingly, the midgap condition $\xi_\alpha=1/2$ diagnoses a
virtual-boundary mode encoded in the photonic band wave functions. It
does not imply nonclassical optical entanglement; it indicates that the
measured linear modes realize the same topological projector as the
corresponding free-fermion band problem.

\subsection{Experimental reconstruction of the entanglement spectrum}
\label{app:ES_experimental_protocol}

The entanglement spectrum is not measured as a direct optical observable.
Experimentally, the required object is the complex single-particle band
projector $P_-(k_f)$, from which the restricted correlation matrix and
its spectrum are obtained by post-processing. The essential experimental
requirement is therefore phase-sensitive reconstruction of the lower-band
projector, rather than an intensity-only measurement.

In the synthetic-frequency platform, the momentum $k_f$ conjugate to the
frequency lattice is mapped to the detection time within one modulation
period $k_f=\Omega t \quad ({\rm mod}\ 2\pi)$. Thus time-resolved coherent transmission spectroscopy gives access to the momentum-resolved response of the synthetic lattice. For the two-sublattice model, the relevant measured quantity is the complex transmission matrix
$S_{\mu\nu}(\omega,k_f),
\mu,\nu=\tilde c,\tilde d$,
where $\nu$ labels the input sublattice channel, $\mu$ labels the output sublattice channel, and $\omega$ is the probe detuning. The measurement must retain both amplitude and phase.
In the weak-probe regime, the complex transmission is governed by the
single-particle Green function,
\begin{equation}
S_{\mu\nu}(\omega,k_f)
=
S^{\rm bg}_{\mu\nu}
+
i\sqrt{\gamma_\mu\gamma_\nu}
\left[
\omega+i\Gamma/2-H(k_f)
\right]^{-1}_{\mu\nu},
\label{eq:app_complex_transmission}
\end{equation}
where $S^{\rm bg}_{\mu\nu}$ is a smooth background, $\gamma_\mu$ is the
external coupling rate, and $\Gamma$ denotes the linewidth. Fitting the
two resonance poles at each $k_f$ gives
\begin{equation}
S_{\mu\nu}(\omega,k_f)
\simeq
S^{\rm bg}_{\mu\nu}
+
\sum_{s=\pm}
\frac{
R^{(s)}_{\mu\nu}(k_f)
}{
\omega-E_s(k_f)+i\Gamma_s/2
}.
\label{eq:app_pole_expansion}
\end{equation}
For a Hermitian two-band Hamiltonian, the residue matrix of the lower
band is proportional to the band projector,
\begin{equation}
R^{(-)}_{\mu\nu}(k_f)
\propto
u_{-,\mu}(k_f)u^{*}_{-,\nu}(k_f).
\label{eq:app_residue_projector}
\end{equation}
After calibration of the external coupling rates, the normalized projector
is therefore reconstructed as
\begin{equation}
P_-(k_f)
=
\frac{
R^{(-)}(k_f)
}{
{\rm Tr}\,R^{(-)}(k_f)
}.
\label{eq:app_projector_from_residue}
\end{equation}
The overall gauge of $|u_-(k_f)\rangle$ cancels in $P_-(k_f)$. The real-space correlation matrix is then obtained from the measured
projector by a discrete Fourier transform,
\begin{equation}
C_{n\mu,m\nu}
=
\frac{1}{N_k}
\sum_{l=0}^{N_k-1}
e^{ik_l(n-m)}
\left[P_-(k_l)\right]_{\mu\nu}.
\label{eq:app_exp_C_fourier}
\end{equation}
A virtual subsystem $A$ is chosen in post-processing, and the restricted
matrix $C^A$ is diagonalized as in Eq.~\eqref{eq:app_CA_diag}. Midgap
levels at $\xi=1/2$ then provide the experimental signature of the
critical topological projector.

In practice, the optical protocol consists of the following steps:
\begin{enumerate}
\item Set the modulation amplitudes and phases to realize the target
effective hopping parameters $J_0,J_1,J_2$.
\item Inject a weak coherent probe into the $\tilde c$ and $\tilde d$
channels separately, and record the phase-sensitive time-resolved output
fields from both channels.
\item Convert the detection time to synthetic momentum using
$k_f=\Omega t$, and obtain the complex matrix
$S_{\mu\nu}(\omega,k_f)$.
\item For each $k_f$, fit the two resonance poles of
$S_{\mu\nu}(\omega,k_f)$ and extract the lower-band residue
$R^{(-)}(k_f)$.
\item Calibrate the external coupling rates, construct
$P_-(k_f)$ from Eq.~\eqref{eq:app_projector_from_residue}, and Fourier
transform it to obtain $C$.
\item Restrict $C$ to the virtual subsystem $A$, diagonalize $C^A$, and identify the entanglement midgap levels at $\xi=1/2$.
\end{enumerate}

\section{Two-dimensional synthetic-frequency extension}
\label{sec:SM_2D_extension}

\subsection{Feasible implementation route}
\label{sec:SM_2D_implementation}

We propose a feasible implementation route for the two-dimensional
extended QWZ model in a programmable synthetic-frequency platform. The
purpose of this discussion is not to provide a fully optimized device
layout, but to show that the required coupling channels can be
assembled from experimentally demonstrated building blocks of
thin-film lithium-niobate (TFLN) frequency synthetic dimensions. In
particular, recent TFLN devices based on two resonators connected by
an electro-optically tunable Mach--Zehnder interferometer (MZI) have
demonstrated controllable same-frequency coupling, cross-frequency
coupling, and ladder Hamiltonians such as the Hall and Creutz
ladders~\cite{wang2025versatile,ma2026reconfigurable}. Such platforms
provide a natural starting point for the present two-leg
frequency-ladder realization.

\begin{figure}
  \centering
  \includegraphics[width=1.0\linewidth]{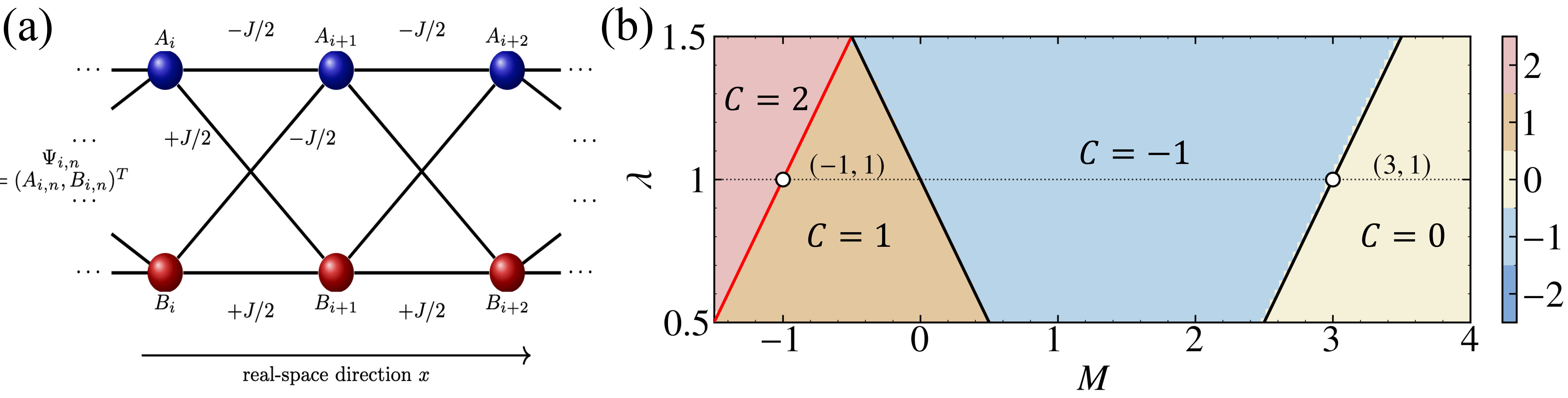} \\
\caption{(a) Real-space coupling schematic for the two-leg implementation of the two-dimensional extended QWZ model. Each cell contains two optical modes, $A_i$ and $B_i$, forming the pseudospin $\Psi_{i,n}=(A_{i,n},B_{i,n})^T$. The four inter-cell links between neighboring cells realize the effective real-space hopping matrix $T_x$. (b) Phase diagram of the two-dimensional extended QWZ model. The color denotes the Chern number $C$ of the lower band. Solid black curves mark the Chern-sector boundaries obtained from the analytical gap-closing conditions. The solid red curve highlights the critical topological phase boundary $M=\lambda-2$, which separates two nontrivial Chern sectors ($C=2$ and $C=1$). The dotted horizontal line indicates the cut $\lambda=1$ used in the main text. The two marked points, $(M,\lambda)=(3,1)$ and $(-1,1)$, correspond respectively to the topological trivial and nontrivial critical points studied in Fig.~4 of the main text.}
\label{fig:supp_QWZ_phase_diagram}
\end{figure}

The extended QWZ Hamiltonian [Eq.~(3) of the main text] can be
written in real space as
\begin{equation}
H = \sum_{i,n} MJ\,\Psi_{i,n}^\dagger\,\sigma_z\,\Psi_{i,n}
+ \sum_{i,n}\!\left[
\Psi_{i+1,n}^\dagger\,T_x\,\Psi_{i,n}
+ \sum_{p=1,2}
\Psi_{i,n+p}^\dagger\,T_f^{(p)}\,\Psi_{i,n}
+ \mathrm{h.c.}\right],
\label{eq:SM_2D_real_space}
\end{equation}
where $\Psi_{i,n}=(A_{i,n},B_{i,n})^T$ is the optical pseudospin
formed by two racetrack resonators in unit cell $i$ at frequency mode $n$ (separated by the free spectral range $\Omega_f$). The hopping matrices are
\begin{equation}
T_x=-\frac{J}{2}\,\sigma_z+\frac{iJ}{2}\,\sigma_y ,
\qquad
T_f^{(p)}=-\frac{J_p}{2}\,\sigma_z+\frac{iJ_p}{2}\,\sigma_x ,
\label{eq:SM_2D_hopping_matrices}
\end{equation}
with $J_1=J$ and $J_2=\lambda J$.

The synthetic-frequency hoppings $T_f^{(p)}$ are implemented using two types of electro-optic modulation. First, the phase modulation on each resonator at frequencies $p\Omega_f$ generates same-ring hopping between frequency modes $n$ and $n+p$. Applying opposite modulation phases on the two resonators yields the diagonal part,
\begin{equation}
A_{i,n}\leftrightarrow A_{i,n+p}:\ -\frac{J_p}{2},
\qquad
B_{i,n}\leftrightarrow B_{i,n+p}:\ +\frac{J_p}{2},
\end{equation}
realizing the $-(J_p/2)\,\sigma_z$ component. Second, an intra-cell MZI connecting $A_i$ and $B_i$ is driven by RF tones that couple modes with both different resonator index and different frequency index. The phase-controlled RF modulation of this MZI generates the couplings $A_{i,n}\leftrightarrow B_{i,n+p}$ and
$B_{i,n}\leftrightarrow A_{i,n+p}$ with a controllable hopping phase; choosing this phase to give the forward hopping $iJ_p\sigma_x/2$ realizes the off-diagonal component of $T_f^{(p)}$. If the two resonators carry a differential detuning $\Delta_{AB}=2MJ$, the cross-frequency tones are shifted from $p\Omega_f$ to
$p\Omega_f\pm\Delta_{AB}$, depending on the direction of the inter-resonator transition.

The real-space hopping $T_x$ (see Fig.~\ref{fig:supp_QWZ_phase_diagram}(a)) requires four inter-cell
couplings per link:
\begin{equation}
\begin{aligned}
&A_i\to A_{i+1}:\ -\tfrac{J}{2},\qquad
&&B_i\to B_{i+1}:\ +\tfrac{J}{2}, \\
&B_i\to A_{i+1}:\ +\tfrac{J}{2},\qquad
&&A_i\to B_{i+1}:\ -\tfrac{J}{2}.
\end{aligned}
\end{equation}
Such a coupling block should be understood as an effective Hamiltonian
element rather than the scattering matrix of a single passive
directional coupler; it can be synthesized by an interferometric
coupling network with calibrated relative phases and
amplitudes. This is the most demanding part of the proposal, but it is
compatible with present programmable MZI-based synthetic-frequency
platforms, where coupling strengths and phases are electro-optically
tunable~\cite{wang2025versatile}.

As a representative parameter regime, one may take
$\Omega_f/2\pi\simeq 9$--$10~\mathrm{GHz}$ and
$J/2\pi\simeq 0.3$--$0.5~\mathrm{GHz}$. For a group index
$n_g\simeq2.2$, this corresponds to a racetrack length
$L=c/(n_g\Omega_f/2\pi)\simeq 14$--$15~\mathrm{mm}$, within the range of existing TFLN frequency-lattice devices. A proof-of-principle device may contain $N_x=4$--$8$ real-space cells and tens of usable frequency modes. Resolving the hopping scale requires the optical linewidth to satisfy $\kappa_{\rm loss}\lesssim J$, while the synthetic-frequency bandwidth should be chosen such that residual dispersion produces onsite-frequency disorder below the relevant topological energy scale. The model can be characterized by frequency-resolved and time-resolved transmission measurements: frequency-resolved output spectra monitor the synthetic-frequency distribution at each spatial output port, while time-resolved spectroscopy along the synthetic-frequency quasi-momentum enables band reconstruction, following established techniques in dynamically modulated frequency lattices. Although a full large-scale implementation requires careful calibration of the inter-cell coupling blocks, RF phases, and resonator dispersion, the required ingredients are available in current programmable synthetic-frequency photonic platforms.

\subsection{Phase diagram of the two-dimensional extended QWZ model}
\label{sec:supp_QWZ_phase_diagram}

To locate the topological sectors of the two-dimensional extended QWZ model, we compute the Chern number of the lower band over the $(M,\lambda)$ plane. The numerical phase diagram is obtained with the Fukui--Hatsugai--Suzuki lattice-gauge method on a discrete Brillouin-zone mesh, while the minimum direct gap is monitored independently to identify the gap-closing lines.

The calculation is performed for $M\in[-1.5,4.5]$ and $\lambda\in[0.5,1.5]$; Fig.~\ref{fig:supp_QWZ_phase_diagram}(b) shows the parameter window relevant to the main text.

The displayed phase boundaries are determined by the simultaneous conditions
$d_x=d_y=d_z=0$.  In the plotted window, the Chern-number-changing boundaries are
\begin{equation}
M=\lambda-2,\qquad
M=1-\lambda,\qquad
M=\lambda+2.
\label{eq:supp_qwz_phase_boundaries}
\end{equation}
These analytical curves agree with the jumps of the numerically computed Chern number.  Along the cut $\lambda=1$, they occur at $M=-1$, $M=0$, and $M=3$.  Among these, the boundary $M=\lambda-2$
(red curve in Fig.~\ref{fig:supp_QWZ_phase_diagram}(b)) is singled out as the critical topological phase boundary: it separates two nontrivial Chern sectors ($C=2$ and $C=1$), and serves as the two-dimensional counterpart of the $\mathcal{W}=1\leftrightarrow2$ critical line of the extended SSH chain discussed above.  The two critical points used in the main text are therefore selected from distinct types of gap closing: at $(3,1)$ the system crosses between a trivial sector ($C=0$) and a Chern sector ($C=-1$), whereas at $(-1,1)$ it crosses between two nontrivial Chern sectors ($C=2$ and $C=1$).  This distinction underlies the different boundary responses discussed in the main text.

\begin{figure}
  \centering
  \includegraphics[width=0.9\linewidth]{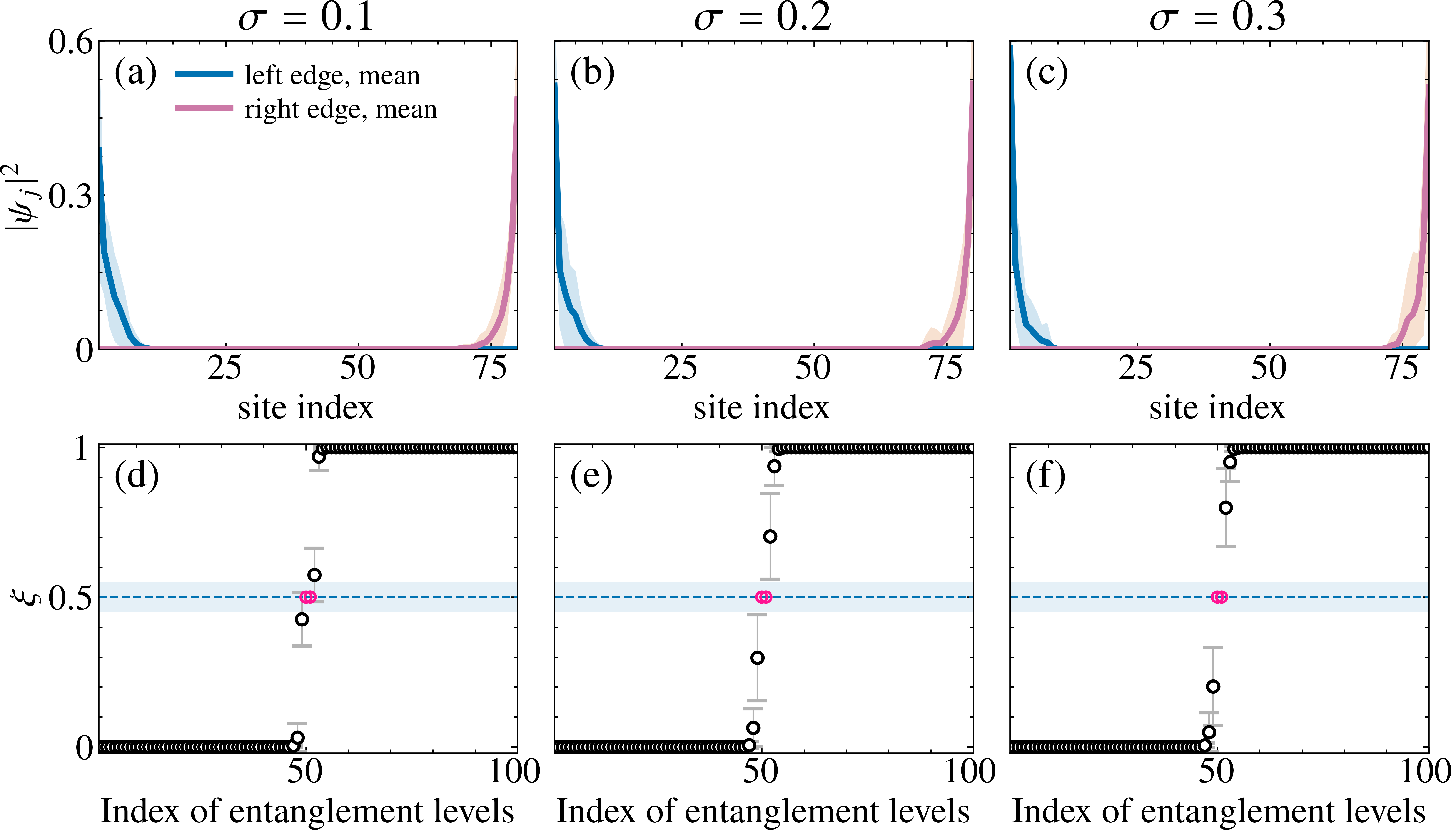}
\caption{Robustness of the 1D critical topological phase against chiral-symmetry-preserving
hopping disorder at $(J_1/J_0,J_2/J_0)=(2.5,1.5)$.  Results are averaged over
$N_{\rm dis}=30$ Gaussian disorder realizations of
Eq.~\eqref{eq:SM_1D_disorder}; shaded bands around the edge states in (a,b,c) denote one standard deviation.
(a)--(c) Disorder-averaged left- and right-edge profiles for
$\sigma=0.1$, $0.2$, and $0.3$ on an open chain with $N_f=80$ unit cells.
(d)--(f) Corresponding half-filled entanglement spectra on an anti-periodic
chain with $N_f=100$ unit cells.  The pair of levels closest to
$\xi=1/2$ remains pinned near the dashed line, confirming the persistence
of the symmetry-protected boundary modes under disorder.
}
\label{figureS3}
\end{figure}

\section{Robustness of critical topological photonic states in 1D and
2D synthetic lattices}
\label{sec:SM_robustness}

A key question is whether the critical topological boundary modes survive realistic imperfections. In this section we test their robustness against symmetry-preserving (structure-preserving) hopping disorder in both the one- and two-dimensional synthetic lattices, using edge-mode profiles and the entanglement spectrum as complementary diagnostics.

\subsection{Robustness against symmetry-preserving perturbations in
1D synthetic lattices}
\label{sec:SM_1D_robustness}

We introduce random hopping fluctuations in the extended SSH chain
while keeping all hoppings inter-sublattice, so that chiral symmetry
is preserved for every disorder realization:
\begin{equation}
J_p(n)=J_p\!\left[1+\sigma\,\eta_p(n)\right],
\qquad
\eta_p(n)\sim\mathcal{N}(0,1),
\qquad p=0,1,2 .
\label{eq:SM_1D_disorder}
\end{equation}
We focus on the representative critical topological point
$(J_0,J_1,J_2)=(1,2.5,1.5)$ and average over $N_{\rm dis}=30$
disorder realizations. Two diagnostics are used: the zero-energy
boundary modes under open boundary conditions, and the half-filled
single-particle entanglement spectrum under anti-periodic boundary
conditions, where symmetry-protected boundary modes appear as
entanglement levels pinned to $\xi=1/2$.

The results are shown in Fig.~\ref{figureS3}. The disorder-averaged edge profiles remain localized at the two ends up to $\sigma=0.3$,
corresponding to $30\%$ relative hopping fluctuations.
Simultaneously, the entanglement spectrum retains a pair of midgap
levels near $\xi=1/2$, with only a small finite-size and
disorder-induced splitting. These observations confirm that the
critical boundary modes are not artifacts of a fine-tuned clean
lattice, but persist under chiral-symmetry-preserving perturbations.

\subsection{Robustness against structure-preserving perturbations in 2D synthetic lattices}
\label{sec:SM_2D_robustness}

We test the robustness of the two-dimensional critical topological
response against structure-preserving imperfections.  We focus on the
nontrivial critical point $(M,\lambda)=(-1,1)$ and introduce random
fluctuations in the hopping amplitudes.  The disorder is chosen to be uniform
along the real-space direction but dependent on the synthetic-frequency index,
so that the momentum $k_x$ remains a good quantum number.  This models
frequency-dependent coupling imperfections while preserving the structure of
the target Hamiltonian.

\begin{figure}
  \centering
  \includegraphics[width=0.9\linewidth]{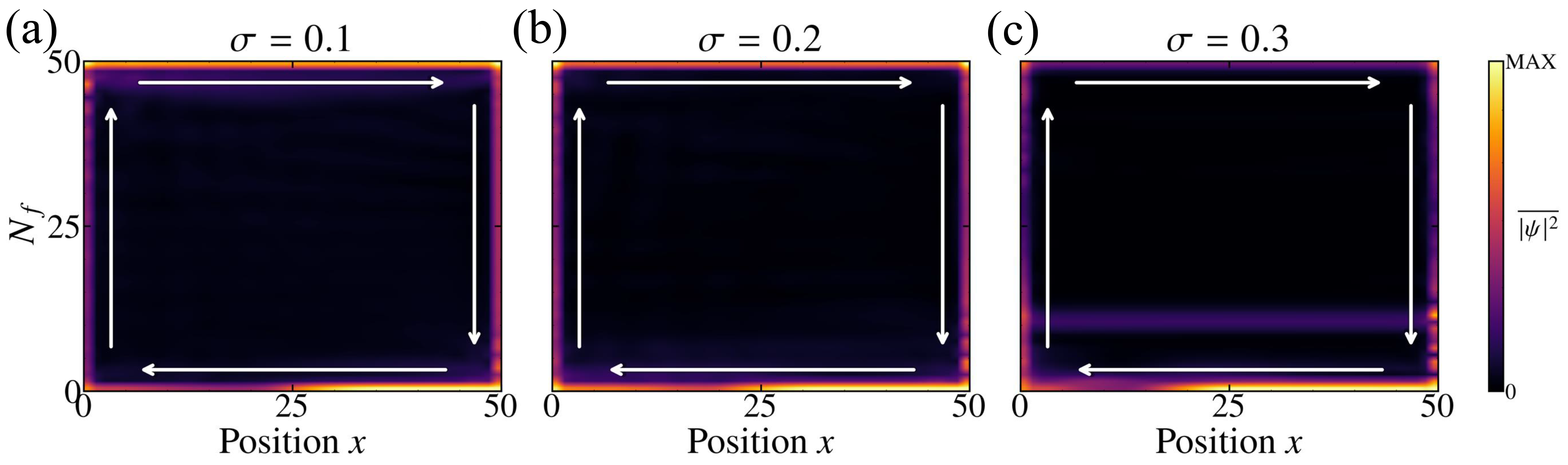}
\caption{
Robustness of the two-dimensional critical topological response against
structure-preserving hopping disorder at $(M,\lambda)=(-1,1)$.
(a)--(c) Time-accumulated intensity of a boundary wavepacket for
$\sigma=0.1$, $0.2$, and $0.3$, respectively.  White arrows indicate the
clockwise boundary-guided propagation.  The persistence edge flow shows that the critical boundary modes remain robust under sizable
hopping-amplitude disorder.
}
\label{figureS4}
\end{figure}

The hopping amplitudes are modulated as
\begin{equation}
T_{\mu}(n_f)\rightarrow
\bigl[1+\sigma \eta_{\mu}(n_f)\bigr]T_{\mu},
\qquad
\mu=x,f,2f ,
\label{eq:SM_2D_hopping_disorder}
\end{equation}
where $\eta_{\mu}(n_f)$ are zero-mean random variables and $\sigma$ controls the relative disorder strength.  No onsite mass disorder or generic Pauli disorder is included, since such terms would change the designed QWZ structure rather than represent coupling imperfections of the synthetic lattice.

Figure~\ref{figureS4} summarizes the results for
$\sigma=0$, $0.1$, and $0.3$.  The panels (a)-(c)  show the time-accumulated intensity of a wavepacket injected near the lower synthetic-frequency edge. The wavepacket remains guided along the boundary even for $\sigma=0.3$ hopping fluctuations, with only moderate disorder-induced leakage.

\end{document}